# A next-generation qPlus-sensor-based AFM setup: Resolving archaeal S-layer protein structures in air and liquid

Theresa Seeholzer[1], Daniela Tarau[2], Lea Hollendonner[1], Andrea Auer[1], Reinhard Rachel[2], Dina Grohmann[2,3], Franz J. Giessibl[1], Alfred J. Weymouth[1]

1) Faculty of Physics, University of Regensburg, 93053 Regensburg, Germany

2) Institute of Microbiology and Archaea Centre, University of Regensburg, 93053 Regensburg, Germany.

3) Regensburg Center for Biochemistry, University of Regensburg, Regensburg, Germany.

Surface-layer (S-layer) proteins form the outermost envelope in many bacteria and most archaea and arrange in 2D quasi-crystalline structures via self-assembly. We investigated S-layer proteins extracted from the archaeon *Pyrobaculum aerophilium* with a qPlus sensor-based atomic force microscope (AFM) in both liquid and ambient conditions and compared it to transmission electron microscopy (TEM) images under vacuum conditions. For AFM scanning, a next-generation liquid cell and a new protocol for creating long and sharp sapphire tips was introduced. Initial AFM images showed only layers of residual detergent molecules (SDS), which are used to isolate the S-layer proteins from the cells. SDS was not visible in the TEM images, requiring a more thorough sample preparation for AFM measurements. These improvements allowed us to resolve the crystal-like structure of the S-layer samples with frequency-modulation AFM in both air and liquid.

**Introduction**

Surface (s)-layers are highly ordered protein structures that surround many bacteria and form the cell wall in almost all archaea, as sketched in Figure 1a[1–4]. S-layers form a robust lattice structure[5,6], working as protective membranes and shielding the cell from potentially extreme environmental conditions. In most cases, S-layers proteins form a monomolecular assembly of identical subunits and maintain their arrangement in 2D structures even after isolation from cells[7]. Lattices in oblique p1 and p2, square p4 or hexagonal p3 and p6 structures (p6 symmetry is depicted in Figure 1b) have been reported in literature[6]. Their robustness and the ability to self-assemble *in vitro* have made S-layers the focus of nanotechnological applications including filters, as building blocks and patterning elements for further self-assembled structures[7–9].

In this work, S-layers from *Pyrobaculum aerophilum* were investigated due to their robust nature and well-known structure[10]. This rod-shaped archaea species was identified in 1993 by Völkl *et al*. after being isolated from a boiling marine water hole in Italy[11]. They are hyperthermophilic organisms and their optimal growth temperature is 100°C at a pH of 7.0. The cell wall structure is composed of S-layer proteins arranged in a p6 lattice[11].

The robustness and the comparably large lattice structure of S-layers make TEM (transmission electron microscopy) a natural imaging method to analyze the symmetrical organisation of the S-layer[12,13]. To resolve the local structure of S-layer structures in natural environments, i.e., in liquid, AFM (atomic force microscopy) studies with no need of staining or vacuum drying, as required for TEM imaging, have provided valuable complementary information: AFM has been used to observe S-layer proteins *in vivo* on bacterial cell walls[14,15], has resolved the self-assembly processes of S-layers[16,17], has visualized the chemical and thermal denaturation processes[18] and has provided high resolution images of S-layers [19]. These previous AFM experiments operated either in static or in dynamic tapping mode (amplitude modulation AFM or AM-AFM). While static mode is rarely used, AM-AFM is the standard method for imaging biological samples as it does not subject samples to lateral forces as the tip scans over the surface.

In this work we present an investigation of the S-layer proteins using frequency-modulation AFM (FM-AFM). FM-AFM is the *de facto* standard for high spatial resolution AFM imaging at low temperature in vacuum. In contrast to AM-AFM, there are two additional active control loops: One to control the excitation frequency so that the tip oscillates at the resonance frequency and the second to control the excitation amplitude to maintain the oscillation at constant amplitude. If the tip is sufficiently close to the surface, the oscillation frequency changes according to the interaction forces between tip and sample atoms, without physical contact between them. Because FM-AFM offers high spatial resolution and has been used to investigate sensitive systems, it is a natural technique to apply to biological samples.

One of the most widely used force sensors for FM-AFM is the qPlus sensor introduced by Giessibl in 1996[20]. Besides applications in ultra-high vacuum and low temperature environments, atomic resolution has been demonstrated in ambient[21–24] as well as in liquid environments[22,25,26].

The qPlus sensor consists of a quartz tuning fork with a tip attached to the end of the oscillating prong. Detection of the qPlus sensor is electric and is based on the piezoelectric effect. To avoid an electrical short, only the tip of the sensor can be immersed in the conducting liquid. Consequently, the first measurements in liquid were made in small drops of water in which the tip was immersed[22,25]. To increase the measurement time, Pürckhauer et al. developed a Teflon-based liquid cell and demonstrated atomic resolution of muscovite mica in several biologically relevant solutions[26]. These first experiments showed the applicability of the qPlus sensor in liquids, but the implementation of stable and long measurement conditions was still challenging. To conquer these existing problems, we developed a new liquid cell design and a new tip preparation method for metal-wire extended, long sapphire tips.

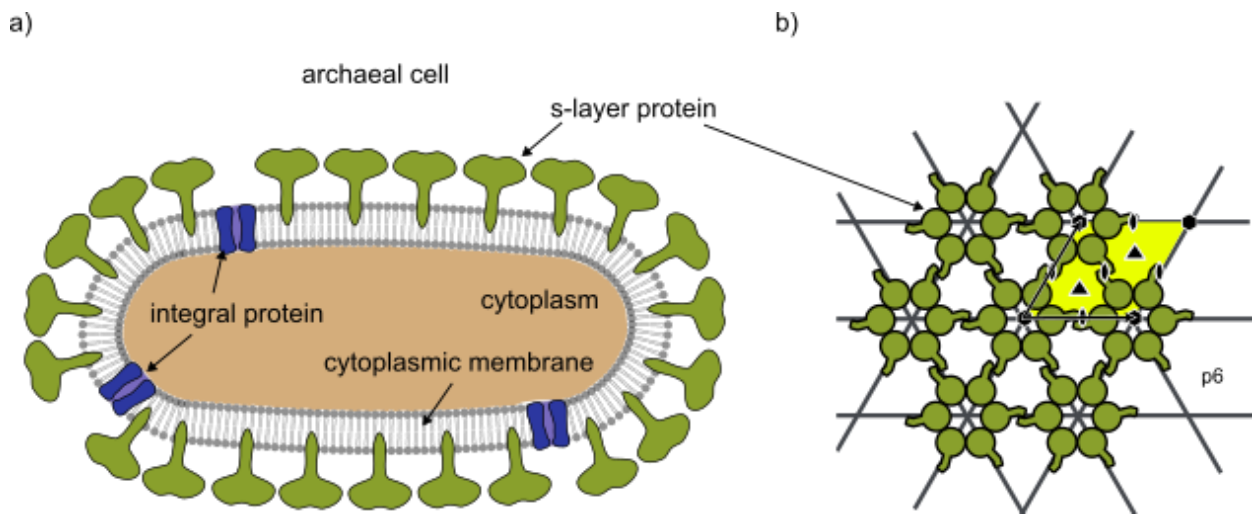

Figure 1. Schematic illustration of S-layer proteins. a) Simplified view of an archaeal cell equipped with an S-layer. The S-layer proteins (green) are directly anchored in the cytoplasmic membrane. Integral membrane proteins (blue) are embedded in the cytoplasmic membrane. b) The S-layer proteins from *Pyrobaculum aerophilum* assemble in a p6 lattice structure.

We present FM-AFM imaging of the S-layer surface structure in an aqueous solution (liquid) as well as in ambient conditions (air) and compare the AFM images to negative stain TEM images acquired in vacuum.

In this paper we highlight the experimental improvements made to a qPlus-based AFM setup to perform reproducible and stable measurements of S-layer protein structures in both air and liquid. Additionally, we optimized the sample preparation for AFM imaging compared to the more widely used TEM method. Negative stain TEM imaging requires a large enough lattice (defined by the cluster size of the staining agent) and the presence of the staining agent. It is therefore not sensitive to residual layers of detergent that is required for the S-layer preparation (in our case sodium dodecyl sulfate, or SDS). However, we observed layers of SDS in the initial AFM images. Thus, additional steps in the sample preparation method were introduced to reduce the SDS concentration and allow successful imaging of the S-layers in a p6 lattice structure using FM-AFM.

## Results and Discussion

**Next generation qPlus-based liquid microscope**

There are several challenges when measuring biological samples in liquid with a qPlus-based AFM. First, as the sensor cannot be immersed fully, the liquid volume is open to air. Consequently, evaporation must be limited to achieve long scanning times (e.g. 2 hours). Related to this, the shape of the meniscus at the water-air interface and the immersion level of the tip must be stable over time since the resonance frequency $f_0$ of the sensor decreases with the immersion depth of the tip[26]. Second, biological samples can be fragile and reactive. Hence, a clean environment and a chemically inert tip material is required. Third, the tip must be long enough to penetrate a reasonable volume of the liquid.

If we first consider the tip material, previous work in ambient and liquid environments have shown great success using sapphire. Sapphire ($Al_2O_3$) is hard compared to other typically used tip materials like tungsten or platinum. In addition, sapphire is chemically inert, acid and base resistant and insoluble in water[27]. Therefore, it is ideal for imaging biological matter. Previously, we produced sapphire tips by splintering bulk crystals[24] but this process leads to a variety of different geometries, shapes and lengths of the splinters (Figure S1). Creating long tips is therefore challenging. To overcome this problem, we developed a procedure to extend short sapphire tips with a metal wire: A small sharp sapphire splinter is glued onto an approximately 2.5 mm long tungsten wire piece with a diameter of 50 µm (see Fig 2 c)). The composite tip with an overall length of around (3.0 - 3.5) mm is then attached to the sensor (see SI for further details of this procedure).

Next, we made improvements to the liquid cell. The first design for a liquid cell proposed by Pürckhauer et al.[26] consisted of Teflon with a rectangular base into which the sample is glued. However, some aspects of this setup were not optimal: It exposed many interfaces between the Teflon and the sample as well as abrupt edges in the corners of the cell. Since Teflon is highly hydrophobic, this resulted in unstable meniscus geometries that changed their shape quickly due to small changes in volume (evaporation) (Figure S2), leading to a jump in the tip immersion depth and consequently in $f_0$. Additionally, the glue of the sample was in direct contact with the solution which is likely to contaminate the sample.

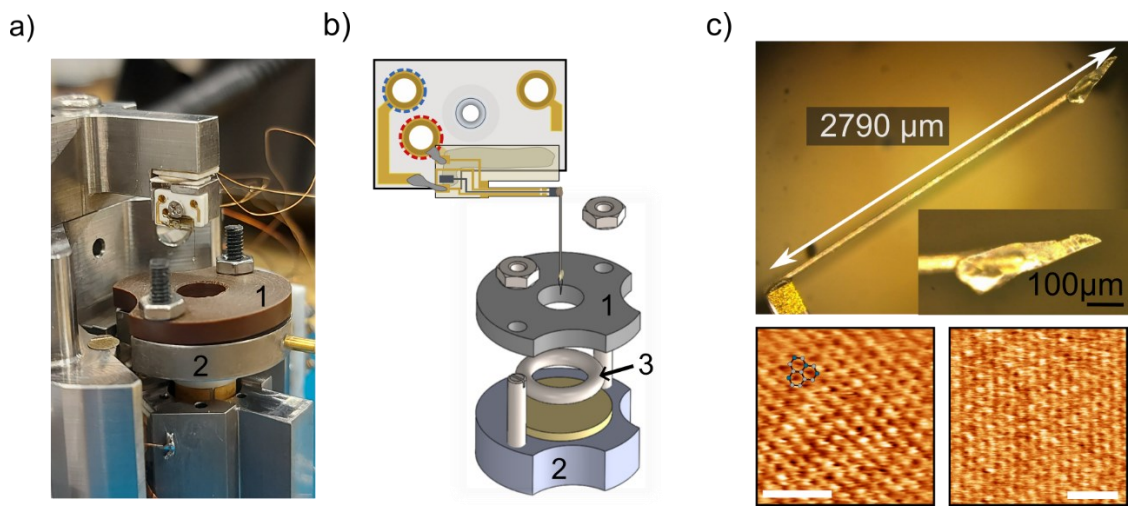

Figure 2. New qPlus based liquid cell: a) Photograph of the new liquid cell design attached at the piezo tube of the microscope. The schematic illustration b) shows the qPlus sensor with a sapphire tip extended via a tungsten wire (shown in more detail in Fig. S3). The cell is composed of three main parts: (1) The top part is made from Tecasint with a cylindrical cut through in the middle serving as the reservoir for the liquid. (2) The base on which the sample is fixed. (3) Using an O-ring, the top part is mechanically pressed onto the sample. The optical micrograph in c) shows the metal-wire extended sapphire tip. To validate the new setup the (001)-plane of the muscovite mica was resolved in both, ambient conditions (left) and in ultrapure water (right). White scale bars are each 2 nm long.

Therefore, we developed a new cell design composed of three main parts (see Fig 2 a)-b)). The first component is the sample holder, part (2), which is made from Kovar™, a Fe-Ni-Co alloy with a low thermal expansion coefficient to minimize thermal drift. At the left- and right-hand side two 1 mm threaded rods are attached. The second part is the top part (1), which has the same basic shape and a height of 2.2 mm. This second part is made of Tecasint® 2011, a polyimide material that is chemically resistant to acids, fats and solvents. A cylindrical cutout in the center with a diameter of 5 mm provides the reservoir for the liquid. As Tecasint® 2011 is slightly hydrophilic, only a small meniscus angle is formed at the water/cell interface. Additional to the middle hole, there are two unthreaded holes to fit the rods attached at the base plate. At the bottom side a groove was drilled for an 8 mm O-ring which works as a seal for the cell. Directly on the sample holder (2), the sample is either fixed using superglue or by mechanical clamping with the top cell part (1). By screwing the nuts on the threaded rods the O-ring is directly pressed onto the sample. Hence, even if glue is used, the liquid contacts only the sample to ensure a clean measurement environment.

To validate the setup, muscovite mica[28] was imaged in liquid as well as in ambient conditions before imaging the S-layer samples. In both environments it was possible to resolve the hexagonal atomic structure of the (001) mica plane, as shown in Fig 2 c).

**Resolving an unexpected SDS-layer**

S-layer proteins were extracted according to a protocol developed for TEM imaging, including a buffer solution containing sodium dodecyl sulfate (SDS) to support cell lysis. For TEM imaging, the S-layer solution was drop casted on a copper grid and negative staining was performed to increase the contrast for TEM imaging. For FM-AFM imaging, similarly, a drop of the solution containing S-layers was deposited on a freshly cleaved mica sheet.

A typical TEM image is shown in Fig. 3a. The proteins arrange in flat sheets with a width of approximately 700 nm and a length of several μm. The p6 arrangement of the single unit cells is also clearly visible. In contrast, the sample surface imaged using AFM, shown in Fig. 3b, appears as having flat terraces with several holes and cracks. The step height of the single holes is (3.4 ± 0.3) nm. On several areas a square lattice with a periodicity of 1.3 nm was observed (see Fig. 3c and d)). In contrast, the TEM images do not show any additional structure aside from the S-layer lattice. As discussed earlier, negative stain TEM imaging requires both the presence of the staining agent and a lattice larger than the staining agent clusters. In the case of uranyl formate (the negative staining agent we used), clusters are on the order of 1.5 nm to 2 nm, which inherently prevents observation of a 1.3 nm lattice.

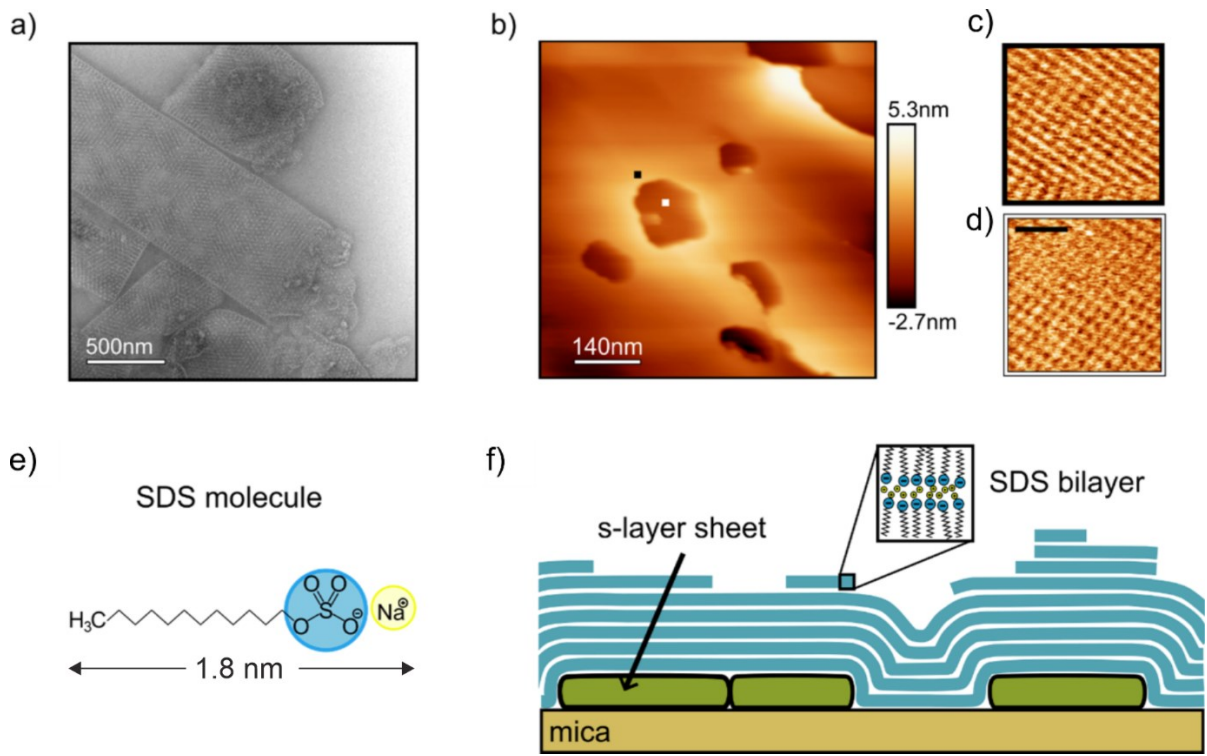

Figure 3. Comparison of the surface structure of S-layer samples prepared according to the original protocol using TEM and FM-AFM. a) TEM image of S-layers adsorbed in flat sheets showing a p6-lattice structure. b) FM-AFM image shows flat terraces with several holes. On different areas of the surface (black- and white-outlined squares) a square lattice c), d) with a periodicity of 1.3 nm is observable, which we propose to be adsorbed SDS-molecules. The black scale bar in d), valid also for c), is a length of 5 nm. e) SDS is an anionic surfactant with a hydrocarbon tail and a polar head-group. Length from Ref.[29]. f) A schematic illustration of a possible surface structure corresponding to b): the S-layer sheets are fully covered by a layer of SDS molecules.

SDS is an anionic surfactant with a hydrocarbon tail and a polar head-group. Previous studies have shown that SDS molecules arrange on different surfaces in stacked bilayers, forming terraces and islands on surfaces[29,30]. Within one bilayer, the negatively charged head groups face each other with the $Na^+$ ions acting as the counter charge (see Fig. 3 e)). The step height of one single SDS bilayer was determined to be 3.7 nm on graphite and $SiO_x$[29]. This led us to conclude that we are imaging stacked bilayers of SDS, which cover the S-layer proteins and/or the mica surface (see Fig. 3f)).

To reduce the concentration of SDS molecules in the sample solution we performed additional purification steps after protein extraction (see Methods and Materials).

Figure 4a shows a TEM image of a sample prepared with the additional washing steps. This image shows little change to the previous TEM image in Figure 3a. The AFM images, on the other hand, showed S-layer proteins arranged in a p6 lattice structure, as can be seen in Figure 4b. Some of the S-layers have lower contrast, as can be seen more clearly in the zoom-in shown in Figure 4c. A line scan of the simultaneously acquired topography is shown in Figure 4d. From the abrupt steps in topography that correspond roughly to the height of SDS bilayers (and multiples thereof) we conclude that SDS bilayers partially cover the S-layers. Therefore, higher contrast in the $\Delta f$ channel is observed when there are no SDS layers upon the S-layers and that when there are one or more bilayers of SDS, the contrast decreases, yet the p6 lattice structure is still visible.

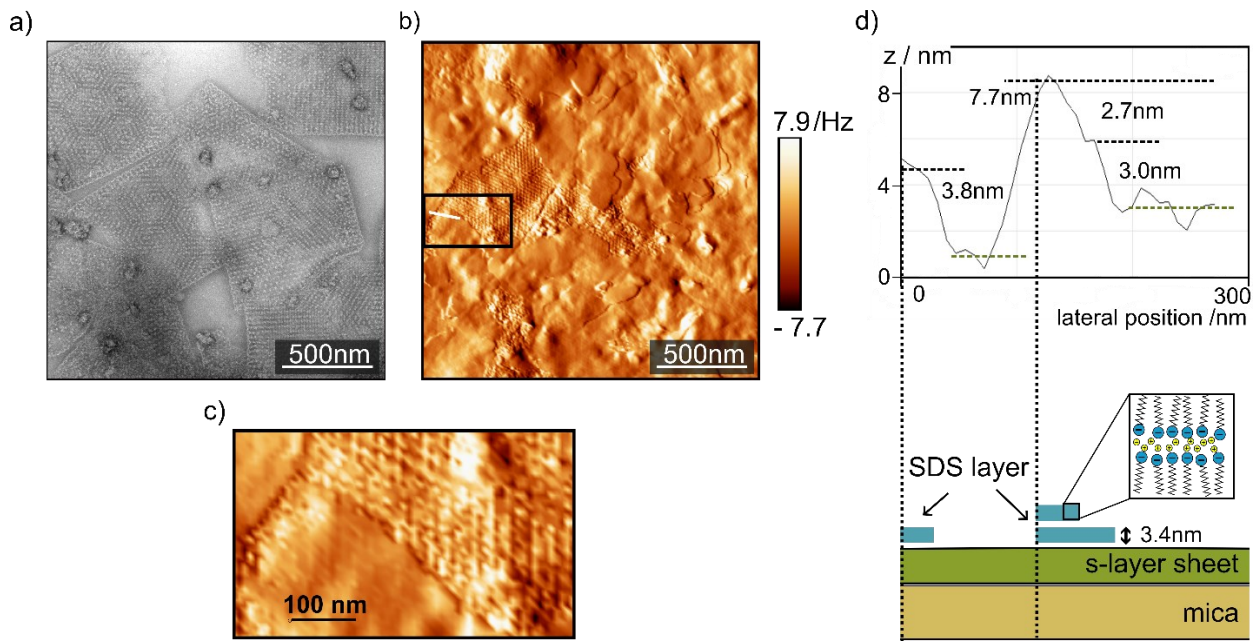

Figure 4. Comparison between TEM and AFM imaging of purified sample solution. a) The TEM image shows flat adsorbed S-layer sheets. b) The AFM images of the dried samples exhibits different surface structures. The S-layer proteins are surrounded by flat layers. The black rectangle corresponds to the zoom in c), demonstrating that the S-layer lattice is still visible with lower contrast through the area of a flat layer. The topographic line d) corresponds to the white line in b) and shows the single step heights between the different surface types (green corresponds to the S-layer lattice, blue to the SDS bilayer). The corresponding surface structure is schematically outlined.

**FM-AFM imaging of S-layer samples in liquid and air**

Using the new liquid cell and an extended sapphire tip, the S-layer proteins adsorbed on the mica surface were imaged with AFM in liquid (ultrapure water) and air.

Figure 5 shows the AFM topography and the corresponding error channels ($\Delta f$ signals). These $\Delta f$ images are effectively spatial derivatives, helpful for observing features when the underlying surface is not perfectly flat.

We compared the S-layer lattice structure in liquid to the structure in air. Figures 5 a) and b) show the z-channel and the corresponding $\Delta f$-channel of the samples in liquid. Figure 5 c) and d) are *z*- and $\Delta f$-channel, respectively) in air.

The areas shown in Fig. 5 reveal a nearly complete coverage of S-layer proteins arranged in a p6-symmetry for both samples in liquid and in air. In the images there are many domains of p6-symmetry with a domain length of approximately 200 nm. To determine the lattice constants within these domains we measured the center-to-center distances along the high symmetry directions, by averaging over the line profiles marked in Fig. 5 a), c). The resulting values are given in Table 1.

An average spacing of (25.3 ± 1.7) nm and (26.0 ± 1.3) nm was found for the S-layer protein structure in air and liquid, respectively.

The lattice distance measured from the AFM data are in excellent agreement with the lattice distance of 25.8 nm extracted from the TEM images shown in Figs 3 and 4. This shows that the lattice structure of the S-layer proteins remains almost the same, independent of the imaging environment.

From the topographic line profiles collected with AFM, the peak-to-peak height of the unit cells was measured resulting in an average height of (2.2 ± 0.3) nm for the samples in air and (2.1 ± 0.7) nm in liquid, where again no significant differences can be observed.

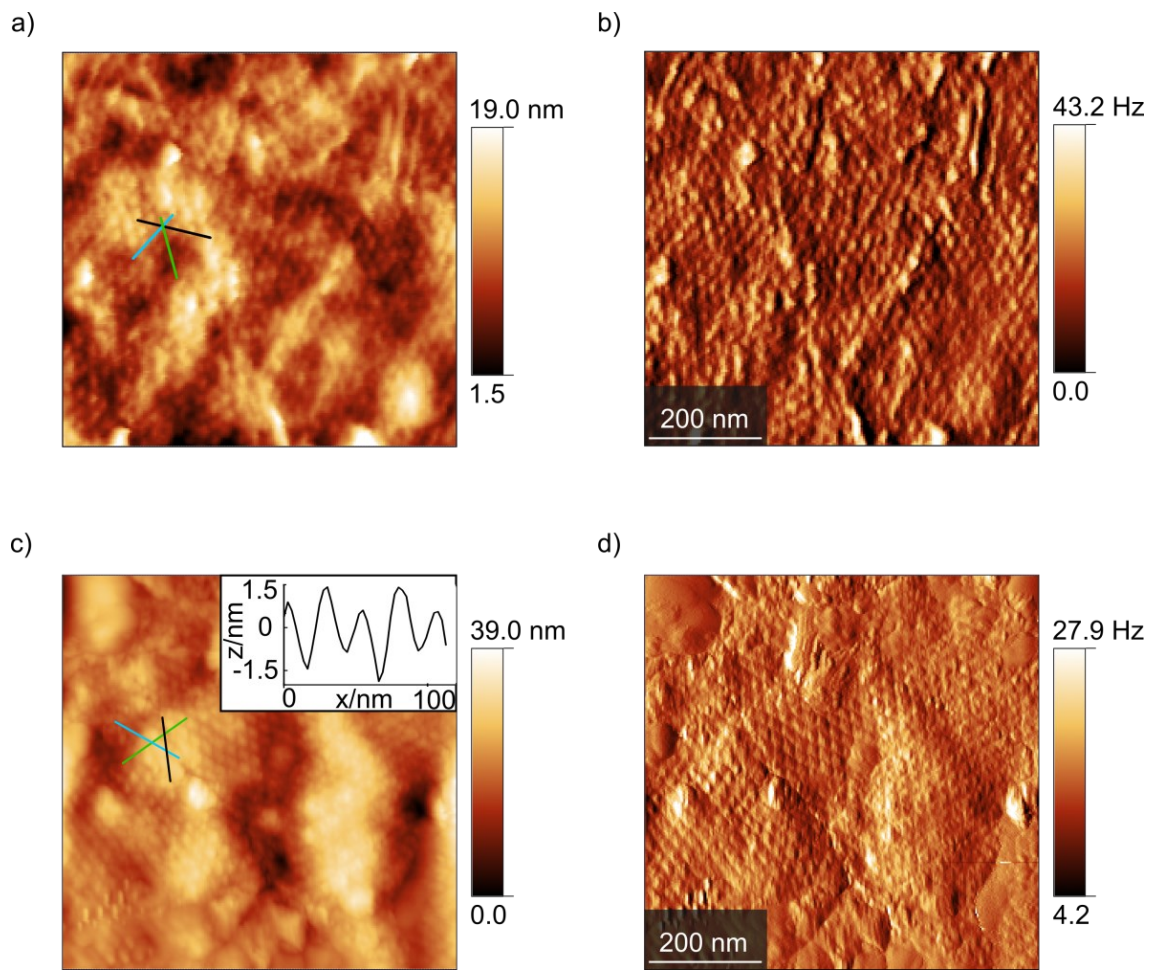

Figure 5. Comparison of the S-layer structure in liquid and in air. The images show two different $(700 \times 700)$ nm$^2$ large areas of the same sample. a) AFM topography and b) the corresponding $\Delta f$-channel of the S-layer proteins in liquid. c) AFM topography and d) the corresponding $\Delta f$-channel of the S-layer proteins in air. The line profiles for analyzing the lattice distances are marked in the topographic images. The inset in c) shows an example of such a line profile.

Table 1 Comparison of the distances (d) and heights (h) between the unit cells in aqueous solution and ambient conditions.

| | green | | blue | | black | | **average** | |
|---|---|---|---|---|---|---|---|---|
| line /nm | d | h | d | h | d | h | **d** | **h** |
| dry | 23.4 | 2.4 | 26.8 | 2.6 | 25.6 | 2.2 | 25.3 ± 1.7 | 2.2 ± 0.3 |
| liquid | 27.5 | 1.8 | 25.0 | 2.0 | 25.4 | 2.5 | 26.0 ± 1.3 | 2.1 ± 0.7 |

Finally, we were able to acquire higher resolution images of the dry sample as shown in Figure 6. These images show that we are sensitive to the internal structure of the monomolecular subunits.

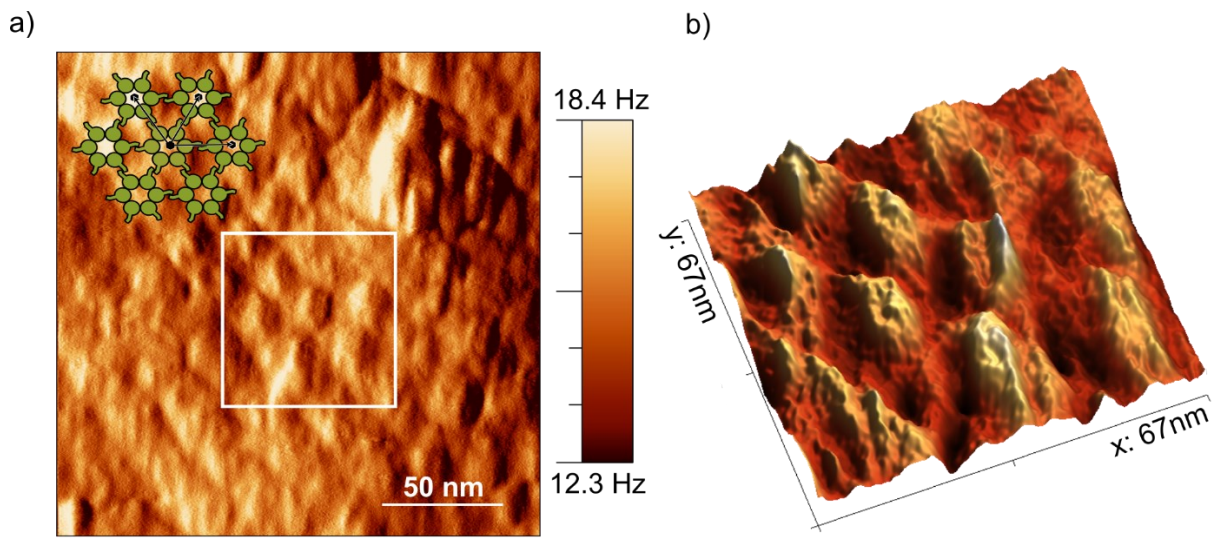

Figure 6. High resolution $\Delta f$ image of the unit cells of the dry S-layer structure. In a) a schematic drawing of the p6 unit cells is included. The arrows indicate the high symmetry directions. The white square of $(67 \times 67)\text{nm}^2$ corresponds to the 3D image shown in b). The elevations of the individual unit cells show a tendency of an internal bump structure.

**Conclusion**

S-layer proteins extracted from the cell wall of *P. aerophilium* were imaged with a qPlus-based FM-AFM in aqueous solution and ambient conditions and compared to TEM images. We presented two key developments which enabled the FM-AFM imaging: First, we introduced a next-generation liquid cell design and a method for creating long sapphire tips. These improvements allow clean and stable measurement conditions for imaging in liquids as well as in ambient conditions over long scanning times. And secondly, we adjusted the S-layer sample preparation from TEM requirements to AFM requirements. While initial AFM images showed only ordered SDS layers, additional filtration steps reduced the SDS concentration and allowed us to successfully image the S-layers using FM-AFM.

With the new microscope setup and the new sample preparation method, we resolved the p6 lattice structure of the S-layer with a periodicity of (25.3 ± 1.7) nm and (26.0 ± 1.3) nm in air and liquid, respectively. The lattice distance is in good agreement with the distance of 25.8 nm measured from TEM images. This shows that the lattice structure of the S-layer proteins remains almost the same even when the imaging environment is changed, as expected from their robustness.

These data furthermore show that FM-AFM can yield structural information on biological systems under biologically relevant conditions achieving spatial resolution sufficient to decipher the structural organization of highly symmetric protein lattices like the S-layer. In contrast to advanced fluorescence-based high resolution imaging techniques like MinFlux[31–33] or DNA PAINT, which provide spatial resolution in the sub-10 nm range, AFM is a label free technique and does not require advanced labeling schemes.

**Methods and Materials**

The AFM images were acquired on a home-built AFM that was previously described[34]. We used a qPlus S0.6 sensor[35] and a sapphire tungsten tip with total tip length of 3.6 mm. For imaging in liquid, the quality factor decreased from 2170 (in air) to 190 (in liquid) and $f_0$ decreased by approximately 200 Hz. AFM images in Fig. 2 and 3b) were taken in quasi-constant height (slow feedback on $\Delta f$) to enhance contrast in the $\Delta f$-channel. AFM images in Figs. 3c) and d), Fig. 4b) as well as Fig. 5 were acquired in the topographic mode, i.e., the height of the tip attempts to follow a constant $\Delta f$ setpoint.

**Sample preparation**

*P. aerophilum* was cultivated as described in Völkl et al[11]. 1 gram *P. aerophilum* cell mass was resuspended by vortexing in 7 ml buffer 1 (20 mM MES/NaOH pH 6.1, 1 mM $MgCl_2$). The cells were disrupted by sonication using a Bandelin Sonopuls HD 2070 for 3 minutes, cycle 5 at a power of 60%. The sonication was repeated 3 times with 1-minute breaks in between. Precipitates originating from the cultivation medium and cell debris were eliminated by short centrifugation at 2,254 g for 7 minutes at 4°C. The supernatant was collected and 1.4 ml SDS-buffer (20 mM MES/NaOH pH 6.1, 1 mM $MgCl_2$, 2% w/v SDS) was added. The mixture was incubated at 80°C for 1 hour to detach the S-layer from the cytoplasmic membrane. After centrifugation at 17,761 g for 2 minutes at 4°C, 3 phases are forming: supernatant, a white pellet and a black pellet. The S-layer is found in the white pellet, which is carefully collected and resuspended in 200-400 µl of SDS-buffer. Samples resulting from this S-layer preparation were used for initial TEM and AFM imaging attempts

To further reduce the concentration of SDS from the S-layer preparation, an optimized purification strategy was developed. To this end the white pellet was not resuspended in SDS-buffer but in buffer 1 (20 mM MES/NaOH pH 6.1, 1 mM $MgCl_2$) instead. In addition, the final S-layer mixture was centrifuged 3 times at 15,871g for 5-10 minutes at room temperature and each time the supernatant was replaced with 200-300 µl fresh buffer 1 in order to diminish the final SDS concentration. This procedure is shown schematically in Fig. S4.

**Negative staining and TEM images acquisition**

The S-layer mixture was applied and blotted on a freshly glow discharged copper grid with 400 mesh from Plano GmbH coated with a continuous carbon layer of around 8 nm thickness. The staining was performed by addition of using uranyl formate solution to the grid and blotting with Whatman paper for 3 times.

The images were acquired at the inhouse 200 kV FEG transmission electron microscope (JEOL Germany GmbH, Freising), using an F416 CMOS camera (TVIPS GmbH, Gauting, Germany). The micrographs were acquired manually using SerialEM (33) at increasing magnifications as shown in figure S5.

ASSOCIATED CONTENT

   **Supporting Information**.

The following files are available free of charge.

Supporting Information (PDF): Includes images of long sapphire tips with sapphire crystal splinters only; the meniscus shape in the Teflon-based cell design; instructions for the construction of the sapphire tips; a schematic of the extraction of the s-layer proteins of *Pyrobaculum aerophilum*; additional TEM images.

AUTHOR INFORMATION

**Corresponding Author**

*University of Regensburg, Regensburg 93053, Germany. orcid.org/0000-0001- 8793-9368;

Email: jay.weymouth@ur.de

**Author Contributions**

AJW, DG and FJG conceived of the experiment. TEM data were collected by DT and RR. AFM data were collected by LH and TS. The new liquid cell was designed by TS and AA. TS, AJW and AA wrote the original draft. All authors contributed to the final version.

**Funding Sources**

ACKNOWLEDGMENT

We would like to thank Korbinian Pürckhauer for support during the initial setup of the AFM experiments. The Grohmann lab would like to thank Katharina Vogl for initial S-layer preparations

ToC Graphic:

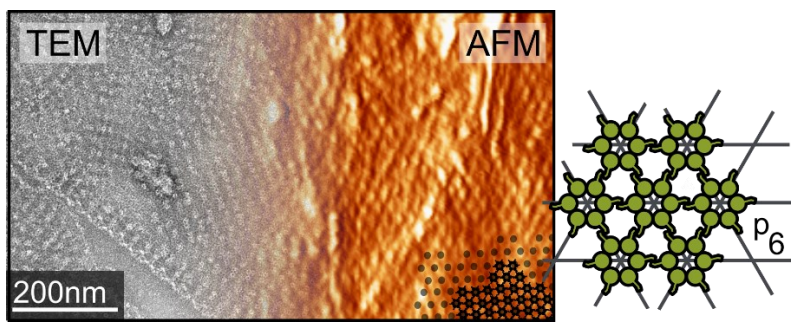

# Supporting Information for

## A next-generation qPlus-sensor-based AFM setup: Resolving archaeal S-layer protein structures in air and liquid

Theresa Seeholzer, Daniela Tarau, Lea Hollendonner, Andrea Auer, Reinhard Rachel, Dina Grohmann, Franz J. Giessibl, Alfred J. Weymouth*

*email: jay.weymouth@ur.de

Includes images of long sapphire tips with sapphire crystal splinters only; the meniscus shape in the Teflon-based cell design; instructions for the construction of the sapphire tips; a schematic of the extraction of the s-layer proteins of *Pyrobaculum aerophilum*; additional TEM images.

**Creating long sapphire tips with sapphire crystal splinters only**

When using only sapphire splinters as tips, we were not able to create sufficiently long tips for successful scanning in the liquid cell (necessary tip length > 2.5 mm).

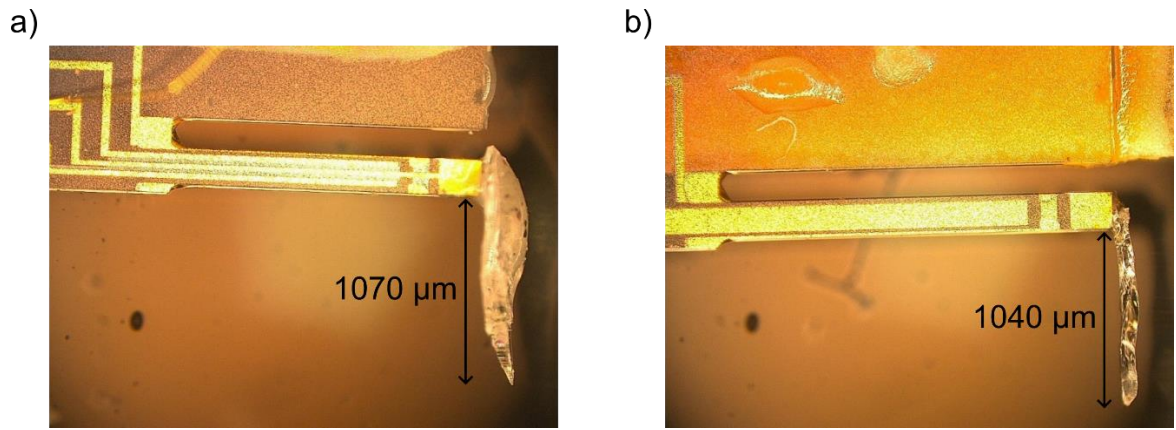

Figure S1. Shown are two qPlus-based sensors with long sapphire splinters as tip. The tip length of a) is 1070 µm and b) is 1040 µm.

**Meniscus shape in the Teflon-based cell design**

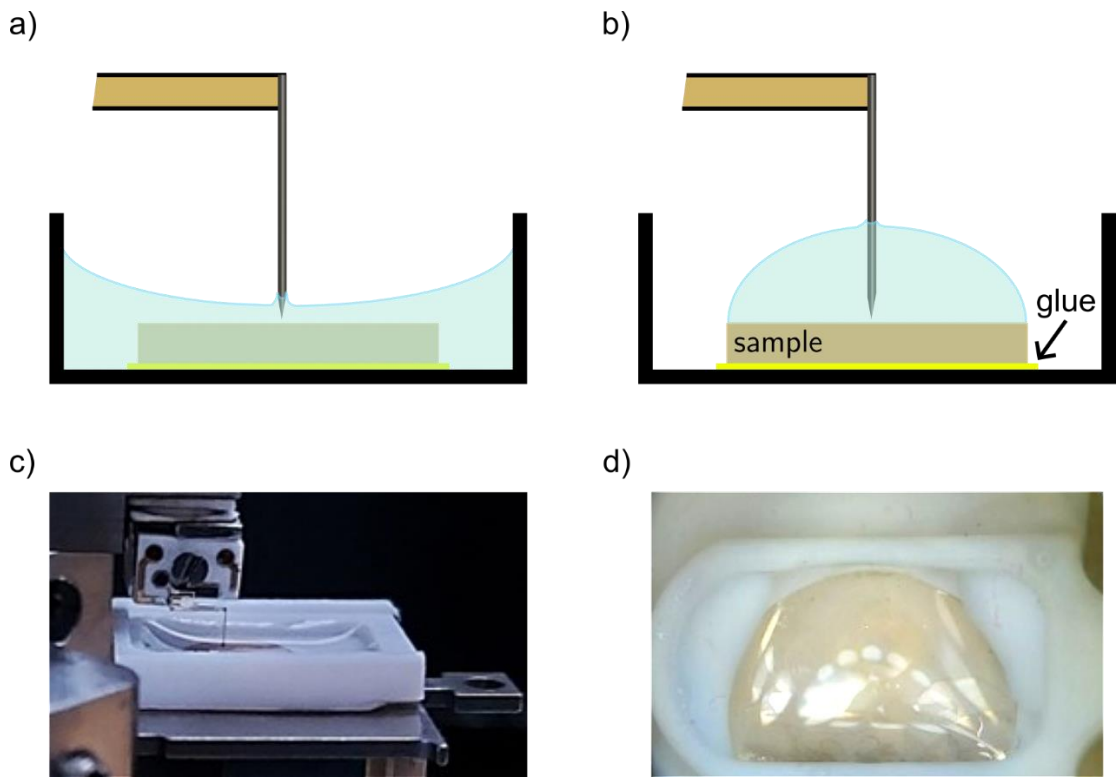

Figure S2. The images show the previous design of a Teflon liquid cell for the qPlus-based system, where the sample is directly attached to the rectangular base of the cell with an adhesive. An example of the meniscus formed due to the cell properties can be seen in a) and b), respectively. Depending on the meniscus formation, the immersion depth of the tip changes significantly. Both meniscus formations have been observed for the same filling of 70 µL in real measurements c)-d). Switching between the configurations occurs in an uncontrolled fashion due to external contact with the system or independently by evaporation processes and therefore leads to unstable measuring conditions.

**Construction of the sapphire tips**

The sapphire tips are assembled by hand under a light microscope. While the wire material does not need to be tungsten (W), we use W as it is very stiff and easy to handle. The wire is cut with a side cutter to the desired length, in our case about 2.5 mm, and placed over a drill hole with 1 mm diameter so that one end is free hanging and the other is fixed by a glass sheet, as shown in Figure S3a. At the free-hanging side, a small droplet of epoxy (H70E) is placed. Using a micro spatula, a small sapphire splinter is picked up. As sapphire is hydrophobic, it usually sticks to the spatula. As depicted in Figure S3a, the splinter is attached at the gluing point of the wire so that the crystal and wire build a straight line. To prevent any unwanted angle between wire and crystal, the splinter should lay flat on the other side of the drill hole. After the epoxy is hardened in the oven, the finished tip is attached to the qPlus sensor, as shown in Figure S3b.

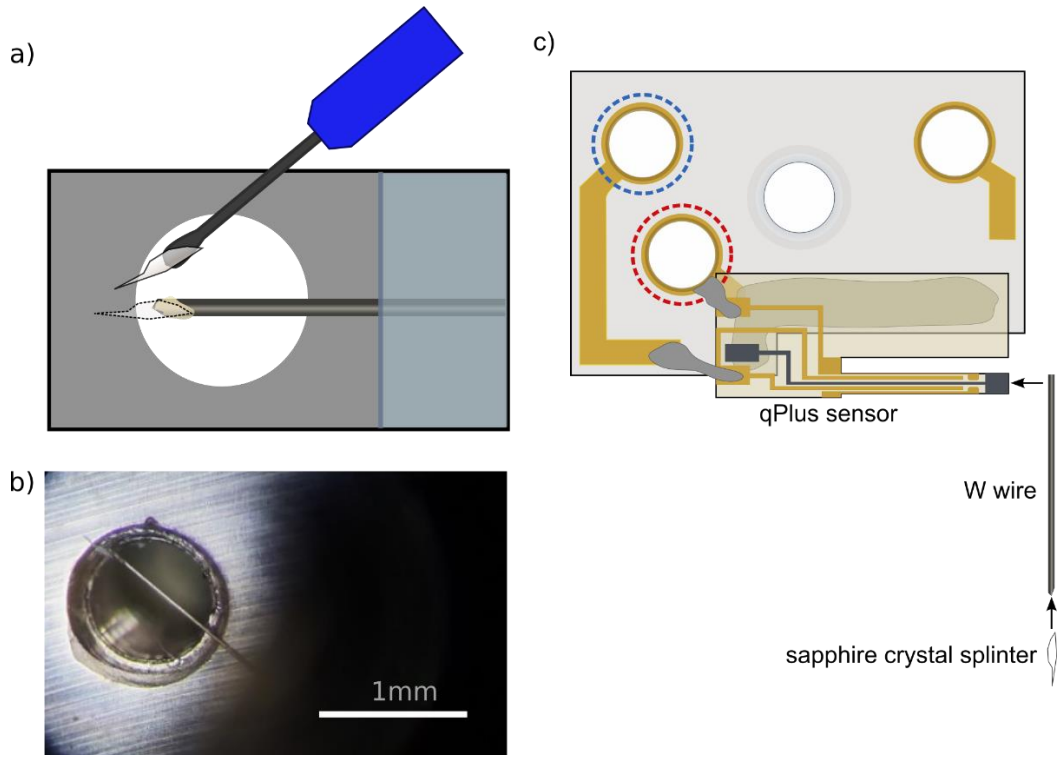

Figure S1. Illustration of the manufacturing of the combined sapphire-tungsten tip. a) Schematic illustration of attaching a crystal splinter to a tungsten wire d = 50 µm using a micro spatula. b) Photograph of the attached splinter at the end of the free-hanging wire. This tip is then attached to the qPlus sensor. c) The qPlus tuning fork is attached to the substrate with epoxy. Then the crystal splinter is glued to the W wire and the W wire is glued to the free tine of the sensor.

**Extraction of the s-layer proteins of *Pyrobaculum aerophilum***

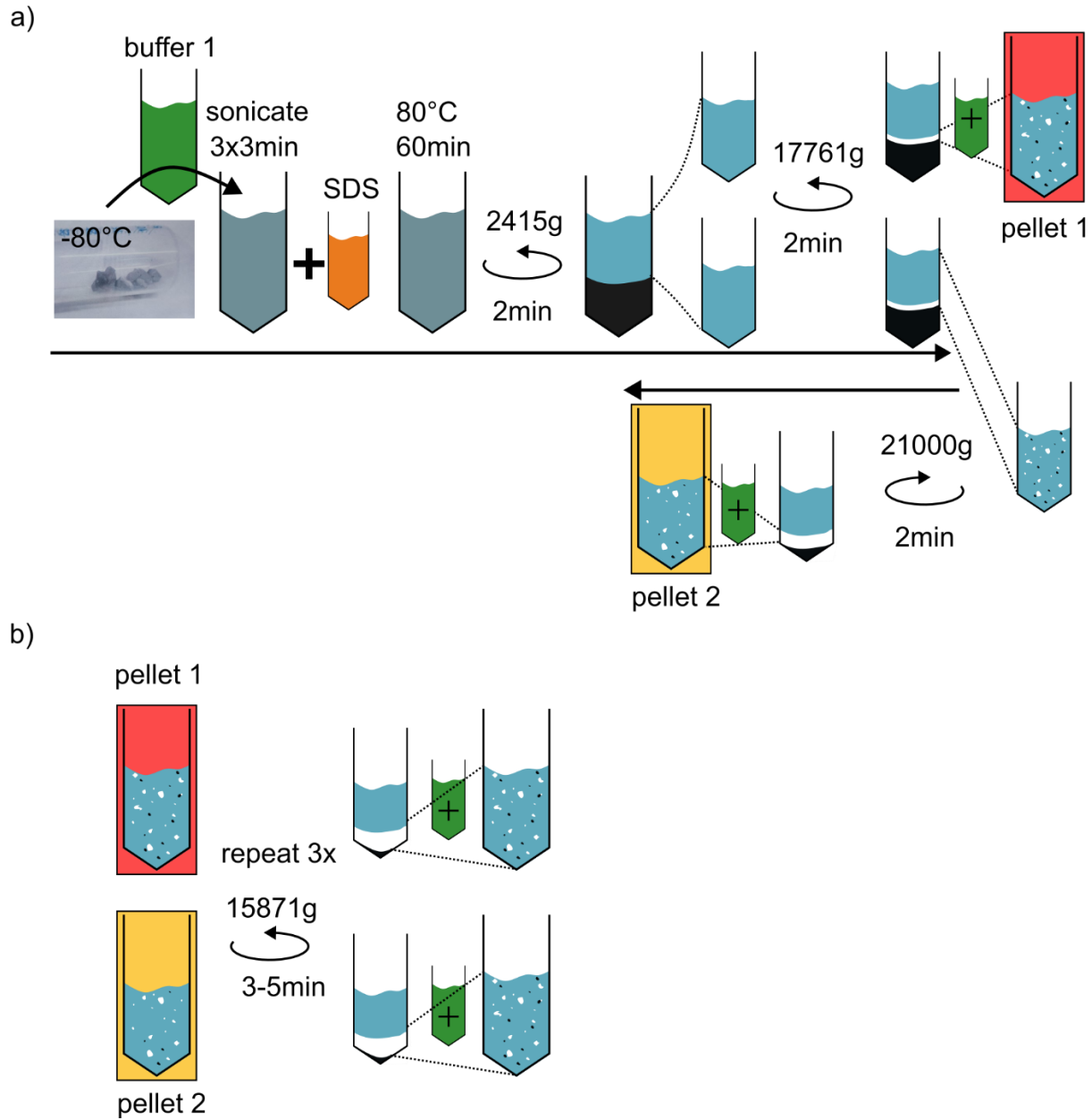

Figure S2. The protocol for the s-layer protein extraction. a) Starting from the left hand with the archaeal cell pellet that were stored at -80°C until used for the S-layer preparation. Shown is a photograph of the storage tube with dark grey cell mass. The circular arrows indicate centrifugation steps. White particles in the supernatant indicate the S-layer proteins. The black parts correspond

to cell debris and further impurities. Pellet 1 (highlighted in red) and pellet (the solution used for AFM imaging, highlighted in yellow) represent the final sample solutions.

**TEM imaging of s-layer proteins of *Pyrobaculum aerophilum***

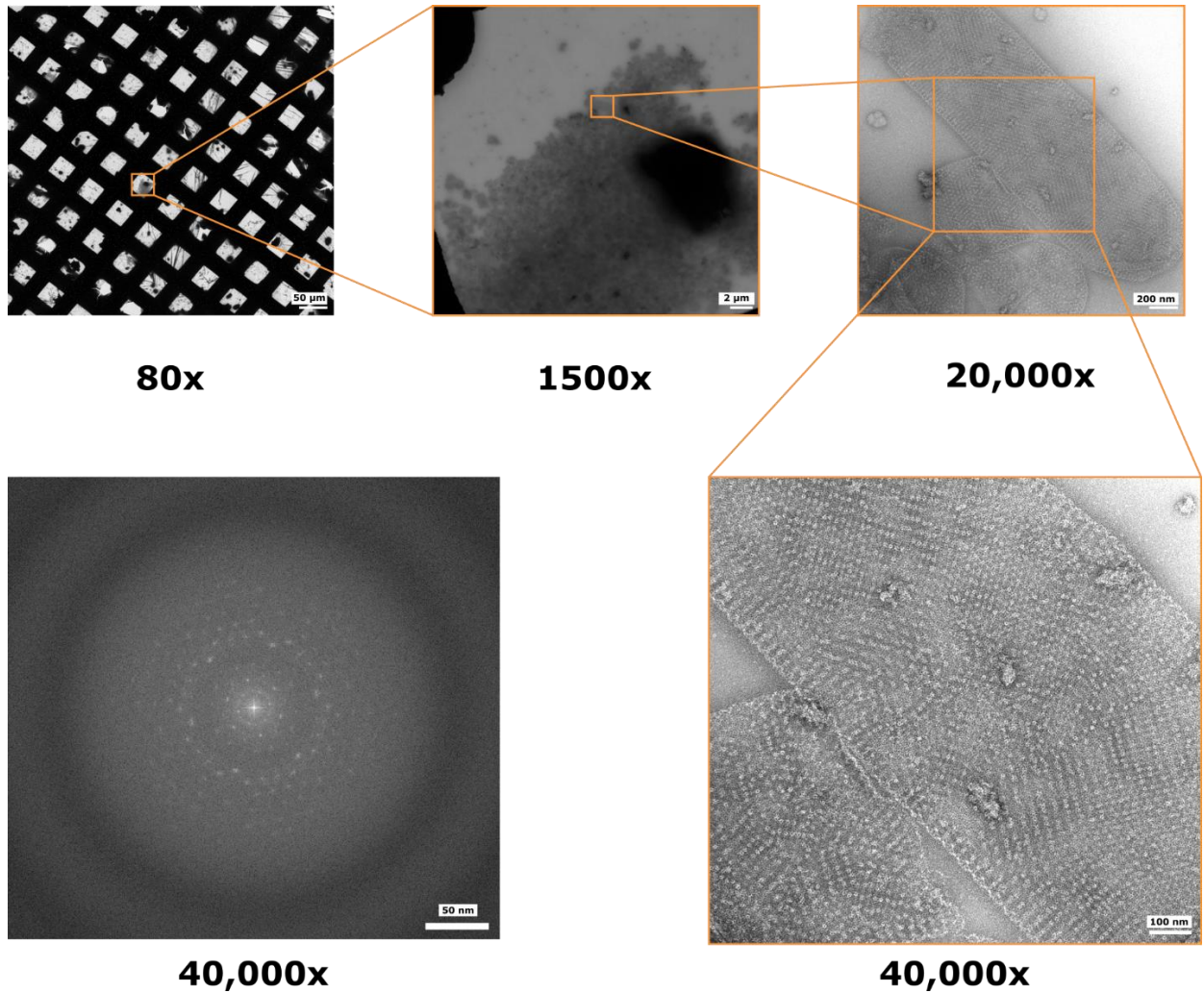

Figure S5. Representation of the general workflow of data acquisition for negative staining samples. Starting with 80x magnification or grid map, following with a square map at 1500x magnification to arrive to the final 40,000x magnification. The power spectrum at 40,000x shows symmetry pattern of the *P. aerophilum* S-layer. Sizes of the scale bars are given in the images.

8